\pdfoutput=1

\documentclass[prd,aps,amsmath,nofootinbib,amssymb,eqsecnum,showkeys,reprint]{revtex4-2}

\usepackage{amsmath, latexsym, amssymb, graphicx, color, multirow, bm, braket, ascmac, bbm, outlines, amsfonts}
\usepackage{hyperref}

\newcommand{\unmarkedfootnotetext}[1]{%
\begingroup
\renewcommand\thefootnote{}%
\footnotetext{#1}%
\addtocounter{footnote}{-1}%
\endgroup
}

\begin{document}

\preprint{APS/123-QED}

\title{Unveiling a Universal Formalism for Quantum Entanglement in Arbitrary Spin Decays}

\author{Junle Pei$^{1,*,\dagger}$, Lina Wu$^{2,*,\ddagger}$, Dianwei Wang$^{3}$, Xiqing Hao$^{3}$, and Tianjun Li$^{3,\S}$}

\affiliation{\vspace{2mm} \\
		$^1$Institute of Physics, Henan Academy of Sciences, Zhengzhou 450046, P. R. China\\
  		$^2$School of Sciences, Xi'an Technological University, Xi'an 710021, P. R. China\\
		$^3$School of Physics, Henan Normal University, Xinxiang 453007, P. R. China}

\date{\today}

\begin{abstract}

We present a comprehensive theoretical framework for probing quantum entanglement in the decay angular distributions of a spin-$S$ particle–antiparticle pair $A\bar{A}$, where each particle decays sequentially into a two-body final state, $A\to B+C$ and $\bar{A}\to\bar{B}+\bar{C}$, with $B(\bar{B})$ carrying spin $b$ and $C(\bar{C})$ being spinless. Starting from the most general polarized initial state, we derive the fully differential angular distribution $\mathcal{W}(\theta_1,\theta_2,\phi_1,\phi_2)$ and identify observables $\langle\cos(2S(\phi_1\mp\phi_2))\rangle$ whose expectation values directly depend on the entanglement-sensitive coefficients $\text{Re}\left(\alpha_{-S,\mp S}\alpha_{S,\pm S}^*\right)$ of the initial state. The proportionality factor $\mathcal{C}(S,b)$ in these relations is computed explicitly. For bosonic decays ($b=0,1,2,\ldots$), $\mathcal{C}(S,b)$ is universal and independent of decay dynamics; in particular, $\mathcal{C}(S,0)=1/2$ for any $S$, and $\mathcal{C}(1,1)=1/8$, matching known results for $W^+W^-$ decays. For fermionic decays ($b=\frac{1}{2},\frac{3}{2},\frac{5}{2}\ldots$), $\mathcal{C}(S,b)$ depends explicitly on the spin analysis powers $\alpha_{A/\bar{A}}$, making entanglement extraction more decay-dependent. We further demonstrate, within the context of $e^+e^-\to\gamma^*\to A\bar{A}$ production, how $\alpha_{A/\bar{A}}$ can be determined experimentally using specific angular observables restricted to the beam-axis region. Our results highlight the special role of bosonic decays in providing clean, model-independent tests of quantum entanglement at colliders, while outlining a pathway for entanglement measurement in fermionic cases through supplementary polarization information.

\end{abstract}

\maketitle
\unmarkedfootnotetext{$^*$These authors contributed equally to this work.}
\unmarkedfootnotetext{$^\dagger$Contact author: peijunle@hnas.ac.cn}
\unmarkedfootnotetext{$^\ddagger$Contact author: wulina@xatu.edu.cn}
\unmarkedfootnotetext{$^\S$Contact author: tli@itp.ac.cn}

\section{Introduction}\label{sec:Intro}

Quantum entanglement, a quintessential feature of quantum mechanics, has evolved from a philosophical conundrum to a vital tool in probing the fundamental laws of nature\cite{PhysRev.47.777,PhysicsPhysiqueFizika.1.195}. In the realm of high-energy physics, the decays of unstable particles produced in collider experiments offer a natural and abundant source of entangled systems~\cite{BESIII:2018cnd,Fabbrichesi:2021npl,Afik:2020onf,Han:2023fci,Cheng:2023qmz,Subba:2024mnl,Subba:2024aut,ATLAS2024,CMS:2024pts,Han:2025ewp,vonKuk:2025kbv,Pei:2025non,Pei:2025yvr,Lin:2025eci,Wu:2025dds,BESIII:2025vsr,Cheng:2025cuv,Pei:2025ito}. The sequential decays of a particle-antiparticle pair, \(A (\to BC)\bar{A}(\to \bar{B}\bar{C})\), provide a rich angular fingerprint from which the quantum correlations—or entanglement—in the initial \(A\bar{A}\) state can be extracted. Analyzing these correlations not only tests quantum mechanics at energy scales never before accessed but also offers a novel, model-independent window into potential new physics that might subtly alter decay dynamics or spin correlations.

The angular distribution of decay products is the primary observable encoding information about the spin state of the parent particles. While the formalism for analyzing spin correlations in specific processes, such as top-quark pair production or the decay of \(W\)-boson pairs, is well-established, a unified treatment applicable to particles and resonances of arbitrary spin \(S\) and decaying to final states of arbitrary spin \(b\) remains less developed. Such a general framework is increasingly necessary as the energy frontier expands, revealing a spectrum of known and potentially new resonances with diverse spin quantum numbers.

The extraction of entanglement observables from angular distributions faces a key dichotomy based on the nature of the final-state particle \(B\). For bosonic decays (\(b=0,1,2,\ldots\)), the relevant coefficients in the angular observables often enjoy a remarkable universality, being independent of the specific decay dynamics. This makes them ideal for clean, direct tests of entanglement. In contrast, for fermionic decays (\(b=\frac{1}{2},\frac{3}{2},\frac{5}{2},\ldots\)), these coefficients typically depend on decay-specific spin analysis powers \(\alpha_{A/\bar{A}}\), introducing a layer of complexity that requires supplementary measurements.

In this work, we bridge this gap by constructing a comprehensive theoretical framework for probing entanglement in the decays of generic spin-\(S\) particle-antiparticle pairs. We derive the complete, most general angular distribution \(\mathcal{W}(\theta_1, \theta_2, \phi_1, \phi_2)\) and identify the observables \(\langle \cos(2S(\phi_1 \mp \phi_2)) \rangle\) that are directly sensitive to the entanglement-sensitive off-diagonal elements (\(\alpha_{-S, \mp S}\alpha^*_{S, \pm S}\)) of the initial \(A\bar{A}\) density matrix. We compute the proportionality factor \(\mathcal{C}(S,b)\) explicitly for both bosonic and fermionic decays, recovering and generalizing known results (e.g., for \(W^+W^-\) and so on).

Furthermore, we address the practical challenge of measuring the spin analysis powers \(\alpha_{A/\bar{A}}\) required for the fermionic case~\cite{ABEL1992304,Li:2024luk,Bechtle:2025ugc,Abel:2025skj,Pei:2025ito}. Focusing on the common production mode \(e^+e^- \to \gamma^* \to A\bar{A}\), we demonstrate how a judicious selection of events along the beam axis, combined with specific angular observables, allows for the direct extraction of \(\alpha_A\alpha_{\bar{A}}\). We provide explicit combinatorial solutions for these observables for a range of half-integer spins \(S\).

This paper is organized as follows. In Section \ref{sec:2}, we establish the general formalism for the angular distribution from the decay of a polarized \(A\bar{A}\) pair. Section \ref{sec:3} presents our central results on entanglement observables, detailing the distinct formulas for the coefficient \(\mathcal{C}(S,b)\) in bosonic (Sec. \ref{sec:3-1}) and fermionic (Sec. \ref{sec:3-2}) decays. Section \ref{sec:4} is devoted to the methodology for extracting spin analysis powers at \(e^+e^-\) colliders. We conclude in Section V with a summary and outlook.

%%%%%%%%%%%%%%%%%%%%%%%%%%%%%%%%%%%%%%%%
\section{GENERAL FRAMEWORK FOR SPIN-DECAY ANGULAR DISTRIBUTIONS} \label{sec:2} 
%%%%%%%%%%%%%%%%%%%%%%%%%%%%%%%%%%%%%%%% 

Consider the decay of the particle–antiparticle pair $A\bar{A}$:
\begin{align}
  A (\to B C) \bar{A}(\to \bar{B} \bar{C})~,
\end{align}
where $A(\bar{A})$ has spin $S$, $B(\bar{B})$ has spin $b$, and $C(\bar{C})$ is spinless (spin $0$). Thus, there must be $S=b,b+1,b+2,...$.
The most general polarization state of $A\bar{A}$ in a pure state can be a quantum superposition of all possible polarization configurations, namely
\begin{align}
    |A\bar{A}\rangle=\sum_{k,j=-S}^S \alpha_{k,j}|k\rangle_A |j\rangle_{\bar{A}}~.
\end{align}
The normalization condition requires
\begin{align}
    \sum_{k,j=-S}^S |\alpha_{k,j}|^2=1~.
\end{align}
Here, $k$ and $j$ are the spin projection quantum numbers of $A$ and $\bar{A}$, respectively, defined along their own momentum directions in the center-of-mass (c.m.) frame.
To set conventions, in the decays $A \to B+C$ and $\bar{A} \to \bar{B}+\bar{C}$ we describe the unit three-momenta of $B$ (in the $A$ rest frame) and $\bar{B}$ (in the $\bar{A}$ rest frame) by spherical angles:
\begin{align}
& \hat{e}_{B/\bar{B}}=(\sin\theta_{1/2}\cos\phi_{1/2},\sin\theta_{1/2}\sin\phi_{1/2},\cos\theta_{1/2})~.    
\end{align}
The polar angles $\theta_i~(i=1,2)$ are measured with respect to the $A$ flight direction $\hat{e}_A$ in the $A\bar{A}$ c.m. frame. The azimuthal angles $\phi_i$ ($\phi_i\in [0,~2\pi]$ for $i=1,2$) are defined relative to an arbitrary axis orthogonal to $\hat{e}_A$ and increase according to the right-hand rule about $\hat{e}_A$.

The angular structures of the decay processes $A \to B+C$ and $\bar{A} \to \bar{B}+\bar{C}$ are governed by the following helicity amplitudes~\cite{Leader:2001nas}:
\begin{align}  
& \langle B,C|k\rangle_A=\sqrt{\frac{2S+1}{{4\pi}}} e^{i(k-\lambda_{1})\phi_1}  
d_{k,\lambda_{1}}^{S}(\theta_1)H_A(\lambda_1)~, \\  
& \langle \bar{B},\bar{C}|j\rangle_{\bar{A}}=\sqrt{\frac{2S+1}{{4\pi}}} e^{i(-j-\lambda_{2})\phi_2}  
d_{\lambda_{2},j}^{S}(\pi-\theta_2)H_{\bar{A}}(\lambda_{2})~.
\end{align} 
Here, \(\lambda_{1}\) and \(\lambda_{2}\) are spin projections of $B$ and $\bar{B}$ defined relative to directions of \(\hat{e}_B\) and \(\hat{e}_{\bar{B}}\), respectively. 
$d_{k,{\lambda}_{1}}^{S}(\theta_{1})$ is the corresponding Wigner $d$-function. Crucially, $H_{A/\bar{A}}(\lambda_{1/2})$ remains independent of both the angular variables ($\theta_{1/2}$ and $\phi_{1/2}$) and the parent particle spin projections $k/j$.
Then, the differential angular distribution of the final state, denoted $\mathcal{W}(\theta_1,\theta_2,\phi_1,\phi_2)$, is~\cite{Pei:2025non,Pei:2025yvr,Pei:2025ito}
\begin{align}
   &\mathcal{W}(\theta_1,\theta_2,\phi_1,\phi_2)
   =\frac{(2S+1)^2}{16\pi^2}\sum_{\lambda_1}h_A(\lambda_{1})\sum_{\lambda_{2}}h_{\bar{A}}(\lambda_{2}) \nonumber\\
   &\sum_{k,j,m,n=-S}^S \alpha_{k,j}\alpha_{m,n}^* 
   e^{i(k-m)\phi_1} e^{i(n-j)\phi_2}\nonumber\\
  & d^{S}_{k,\lambda_1}(\theta_1)
   d^{S}_{m,\lambda_1}(\theta_1)
   d^{S}_{\lambda_{2},j}(\pi-\theta_2)
   d^{S}_{\lambda_{2},n}(\pi-\theta_2) \label{zzd}
\end{align}
with
\begin{align}
   h_{A/\bar{A}}(\lambda_{1/2})=\frac{\left|H_{A/\bar{A}}(\lambda_{1/2})\right|^2}{\sum_{\lambda^\prime_{1/2}} \left|H_{A/\bar{A}}(\lambda^\prime_{1/2})\right|^2}~.
\end{align}

In this work, for simplicity, we assume that both $B$ and $\bar{B}$ possess only two helicity configurations, i.e., $\lambda_{1/2}=\pm b$. Thus, the spin analysis powers are defined by
\begin{align}
\alpha_{A/\bar{A}}=\frac{\left|H_{A/\bar{A}}(-b)\right|^2-\left|H_{A/\bar{A}}(b)\right|^2}{\left|H_{A/\bar{A}}(-b)\right|^2+\left|H_{A/\bar{A}}(b)\right|^2}~.
\end{align}
Then, $\mathcal{W}(\theta_1,\theta_2,\phi_1,\phi_2)$ becomes
\begin{align}
    & \mathcal{W}(\theta_1,\theta_2,\phi_1,\phi_2) \nonumber\\ 
        =&\frac{(2S+1)^2}{64\pi^2}\sum_{\lambda_1,\lambda_{2}=\pm b}\left(1-{\rm Sign}[\lambda_1]\alpha_A\right)\left(1-{\rm Sign}[\lambda_{2}]\alpha_{\bar{A}}\right) \nonumber\\ 
        &\sum_{k,j,m,n=-S}^S \alpha_{k,j}\alpha_{m,n}^* e^{i(k-m)\phi_1} e^{i(n-j)\phi_2} \nonumber\\ 
        & d^{S}_{k,\lambda_1}(\theta_1)
        d^{S}_{m,\lambda_1}(\theta_1)
        d^{S}_{\lambda_{2},j}(\pi-\theta_2)
        d^{S}_{\lambda_{2},n}(\pi-\theta_2)~,
\end{align}
where $\text{Sign}[\pm b]=\pm 1$.

%%%%%%%%%%%%%%%%%%%%%%%%%%%%%%%%%%%%%%%%
\section{ENTANGLEMENT OBSERVABLES FOR ARBITRARY SPINS} \label{sec:3} 
%%%%%%%%%%%%%%%%%%%%%%%%%%%%%%%%%%%%%%%% 

The presence of quantum entanglement in \(A\bar{A}\) pairs can be probed by analyzing the statistical averages of observables associated with the decay processes \(A\to B+C\) and \(\bar{A}\to \bar{B}+\bar{C}\). Specifically, for arbitrary particle spins \(S\) and \(b\), the following formal relations hold:
\begin{align}
    & \langle \cos (2S(\phi_1-\phi_2)) \rangle=  \mathcal{C}(S,b) \left(\alpha_{-S,-S}\alpha^*_{S,S}+\alpha^*_{-S,-S}\alpha_{S,S}\right)~, \label{eq:cos1}\\
    & \langle \cos (2S(\phi_1+\phi_2)) \rangle= \mathcal{C}(S,b) \left(\alpha_{-S,S}\alpha^*_{S,-S}+\alpha^*_{-S,S}\alpha_{S,-S}\right)~.\label{eq:cos2}
\end{align}
The explicit expression for the coefficient $\mathcal{C}(S,b)$ is process dependent and is derived in Appendix~\ref{app1}.

\subsection{Bosonic decays}\label{sec:3-1}

For the case where $B$ is a boson ($b=0,1,2,...$), the coefficient $\mathcal{C}(S,b)$ in Eq.\eqref{eq:cos1}-\eqref{eq:cos2} takes the form
\begin{equation}
   \mathcal{C}(S,b)=\frac{1}{2}\frac{\Gamma^4(S+1)}{\Gamma^2(S-b+1)\Gamma^2(S+b+1)}~.  \label{eq:Cboson}
\end{equation}
A notable feature of this coefficient is that it is independent of the spin analysis powers $\alpha_{A/\bar{A}}$.
Moreover, in the limit of large spin $S\to \infty$, the coefficient approaches $C(S,b)\to \frac{1}{2}$, which is independent of the final-state spin $b$.

In particular, when the final-state particle $B$ is a (pseudo-)scalar ($b=0$), although in this case $B~(\bar{B})$ has only a single helicity configuration, we can still employ Eq.~(\ref{zzd}) to obtain the angular distribution of the final-state particles. We find that, in this situation,
\begin{equation}
    \mathcal{C}(S,0)=\frac{1}{2}~,~~(\rm for~any~S\geq 0)~,
\end{equation}
which also satisfies Eq.~\eqref{eq:Cboson} and is completely independent of both $S$ and the decay dynamics.  This universal result applies to hadronic decays $A\to B+C$ such as 
$K^*(892)\to K\pi$, $\rho(770)\to \pi\pi$, $\Psi(2S)\to \pi\pi$ and so on.

Moreover, for \(S=b=1\), we obtain
\begin{equation}
    \mathcal{C}(1,1)=\frac{1}{8}\;,
\end{equation}
which agrees with the result given in the Ref.~\cite{Pei:2025non} for \(W^-(\to e^- \bar{\nu}_e)W^+(\to e^+ {\nu}_e)\)\footnote{Due to different choices of reference frames, one actually has \(\phi_{e^-}=\phi_1\) and  \(\phi_{e^+}=-\phi_2\).}:
\begin{equation}
    \cos(2(\phi_{e^-}-\phi_{e^+}))=\frac{1}{8}\left(\alpha_{-1,1}\alpha_{1,-1}^*+\alpha_{-1,1}^*\alpha_{1,-1}\right).
\end{equation}
This consistency holds because, when the masses of both the electron and the electron neutrino are neglected, only configurations with \(\lambda_{e^-/e^+}=\frac{1}{2}\) and \(\lambda_{\bar{\nu}_e/{\nu}_e}=-\frac{1}{2}\) or \(\lambda_{e^-/e^+}=-\frac{1}{2}\) and \(\lambda_{\bar{\nu}_e/{\nu}_e}=\frac{1}{2}\) are allowed, i.e. \(\lambda_{e^-/e^+}-\lambda_{\bar{\nu}_e/{\nu}_e}=\pm 1\). This is essentially the same situation as $b=1$ and \(\lambda_{1/2}=\pm 1\).
Since the photon possesses only two transverse polarization states, analogous processes include \(K^{*}(892)\to \gamma+K^\pm\), \(D_s^{*+}\to \gamma+D_s^{+}\) and similar decays.

\subsection{Fermionic decays}\label{sec:3-2}

If $B$ is a fermion ($b=\frac{1}{2},\frac{3}{2},\frac{5}{2},...$), the coefficient $\mathcal{C}(S,b)$ in Eq.\eqref{eq:cos1}-\eqref{eq:cos2} takes the form
\begin{equation}
   \mathcal{C}(S,b)=-\frac{1}{2}\alpha_A\alpha_{\bar{A}}\frac{\Gamma^4(S+1)}{\Gamma^2(S-b+1)\Gamma^2(S+b+1)}~. \label{Eq:Cfermion}  
\end{equation}
Unlike the bosonic cases, this expression depends explicitly on the spin analysis powers $\alpha_{A/\bar{A}}$. Consequently, extracting the observable generally requires independent knowledge or measurement of $\alpha_{A/\bar{A}}$.

The most common situation corresponds to $b=\frac{1}{2}$, as in many weak or electromagnetic decays of heavy hadrons. For instance, in the process
$\Lambda\to p+\pi^-$, the coefficient in Eq. \eqref{Eq:Cfermion} is~\cite{Pei:2025yvr} 
\begin{equation}
    \begin{split}
        \mathcal{C}(\frac{1}{2},\frac{1}{2})=-\frac{1}{32}\pi^2\alpha_\Lambda\alpha_{\bar{\Lambda}} ~.
    \end{split}
\end{equation}

%%%%%%%%%%%%%%%%%%%%%%%%%%%%%%%%%%%%%%%%
\section{EXTRACTING SPIN ANALYSIS POWERS AT $e^+e^-$ COLLIDERS} \label{sec:4} 
%%%%%%%%%%%%%%%%%%%%%%%%%%%%%%%%%%%%%%%% 

Because of the explicit $\alpha_A\alpha_{\bar A}$ dependence, entanglement observables for fermionic $B$ and $A$ are more sensitive to the details of the decay dynamics than their bosonic counterparts. This makes the bosonic case especially attractive for clean, model‑independent tests of entanglement at colliders.

To probe quantum entanglement in the fermionic case ($b=\frac{1}{2},\frac{3}{2},\frac{5}{2},...$), it is often necessary to extract $\alpha_{A/\bar{A}}$ using dedicated observables together with additional polarization information~\cite{ABEL1992304,Li:2024luk,Bechtle:2025ugc,Abel:2025skj,Pei:2025ito}. In the following, focusing on $A\bar{A}$ pairs produced at electron–positron colliders via $e^+ e^-\to \gamma^* \to  A \bar{A}$, we demonstrate how to determine $\alpha_{A/\bar{A}}$.
For the $e^+ e^-\to \gamma^* \to  A \bar{A}$ process, since the intermediate propagator is a virtual photon, in the zero electron mass limit, for events where $A~(\bar{A})$ is emitted along the beam direction ($\cos \theta_A=0~\text{or}~\pi$), the allowed polarization states are restricted to $|k\rangle_A |k+1\rangle_{\bar{A}}~(k=-S,-S+1,...,S-1)$ or $|k+1\rangle_A |k\rangle_{\bar{A}}$, as well as quantum superpositions of these configurations. 
Therefore, when we restrict our attention to events with $\cos \theta_A=0~\text{or}~\pi$, we obtain
\begin{align}
    \sum_{k=-S}^{S-1}\left( \left| \alpha_{k,k+1} \right|^2+\left| \alpha_{k+1,k} \right|^2\right)=1~,~\alpha_{k,j\ne k\pm1}=0~. 
\end{align}

For an angular observable $\mathcal{O}(\theta_1,\theta_2)$ that depends only on $\theta_1$ and $\theta_2$, its statistical average is given by
\begin{align}
    \langle \mathcal{O}(\theta_1,\theta_2) \rangle=\sum_{k,j=-S}^S \mathcal{O}_{k,j}|\alpha_{k,j}|^2~.
\end{align}
Assume that $\mathcal{O}(\theta_1,\theta_2)$ further satisfies
\begin{align}
\langle\mathcal{O}(\theta_1,\theta_2)\rangle=&\alpha_A\alpha_{\bar{A}}\mathcal{D}(S,b)\sum_{k=-S}^{S-1}\left( \left| \alpha_{k,k+1} \right|^2+\left| \alpha_{k+1,k} \right|^2\right) \nonumber\\
&+\text{other terms}~,\label{sz}
\end{align}
where “other terms” denotes contributions associated with $|\alpha_{k,j}|^2$ for which $|k-j|\ne 1$.
For such angular observable $\mathcal{O}(\theta_1,\theta_2)$, when restricting to $A\bar{A}$ events with $\cos \theta_A=0~\text{or}~\pi$, we denote its statistical average by $\langle\mathcal{O}(\theta_1,\theta_2)\rangle_{\text{beam}}$, which is given by
\begin{align}
\langle\mathcal{O}(\theta_1,\theta_2)\rangle_{\text{beam}}=\alpha_A\alpha_{\bar{A}}\mathcal{D}(S,b)~.
\end{align}

In what follows, for $b=\frac{1}{2}$ and $S\in \{\frac{1}{2},\frac{3}{2},\frac{5}{2},\frac{7}{2},\frac{9}{2}\}$ we provide solutions for $\mathcal{O}(\theta_1,\theta_2)$ that satisfies Eq.~(\ref{sz}); we emphasize that such solutions are not unique.

\begin{itemize}
    \item For $S=\frac{1}{2}$,
    \begin{align}
        \langle \cos\theta_1 \cos\theta_2 \rangle_{\text{beam}}=\frac{1}{9}\alpha_A \alpha_{\bar{A}}~.
    \end{align}  
    \item For $S=\frac{3}{2}$,
    \begin{align}
       & \langle \cos\theta_1 \cos\theta_2 \rangle_{\text{beam}}\nonumber\\
       &-\frac{1}{6}\langle \cos(3\theta_1) \cos\theta_2 +\cos\theta_1 \cos(3\theta_2) \rangle_{\text{beam}}\nonumber\\
       &=-\frac{2}{525}\alpha_A \alpha_{\bar{A}}~.
    \end{align}
    \item For $S=\frac{5}{2}$,
    \begin{align}
       & \langle \cos\theta_1 \cos\theta_2 \rangle_{\text{beam}}\nonumber\\
       &-\frac{18}{31}\langle \cos(3\theta_1) \cos\theta_2 + \cos\theta_1 \cos(3\theta_2) \rangle_{\text{beam}}\nonumber\\
       &+\frac{9}{31}\langle \cos(3\theta_1) \cos(3\theta_2)\rangle_{\text{beam}} \nonumber\\
       &=\frac{4}{7595}\alpha_A \alpha_{\bar{A}} ~.
    \end{align}
    \item For $S=\frac{7}{2}$,
    \begin{align}
       & \langle \cos\theta_1 \cos\theta_2 \rangle_{\text{beam}}-\frac{72479}{192141}\langle \cos(3\theta_1) \cos(3\theta_2)\rangle_{\text{beam}}\nonumber\\
       &+\frac{8437}{21349}\langle \cos(5\theta_1) \cos\theta_2 + \cos\theta_1 \cos(5\theta_2) \rangle_{\text{beam}} \nonumber\\
       & -\frac{6292}{106745}\langle \cos(5\theta_1) \cos(3\theta_2) + \cos(3\theta_1) \cos(5\theta_2) \rangle_{\text{beam}} \nonumber\\
       &=-\frac{32}{448329}\alpha_A \alpha_{\bar{A}}~.
    \end{align}
     \item For $S=\frac{9}{2}$,
    \begin{align}
       & \langle \cos\theta_1 \cos\theta_2 \rangle_{\text{beam}}-\frac{1297751}{3854493}\langle \cos(3\theta_1) \cos(3\theta_2)\rangle_{\text{beam}}\nonumber\\
       &+\frac{44135}{142759}\langle \cos(5\theta_1) \cos\theta_2 + \cos\theta_1 \cos(5\theta_2) \rangle_{\text{beam}} \nonumber\\
       & -\frac{23660}{428277}\langle \cos(5\theta_1) \cos(3\theta_2) + \cos(3\theta_1) \cos(5\theta_2) \rangle_{\text{beam}} \nonumber\\
       & +\frac{9464}{428277}\langle \cos(5\theta_1) \cos(5\theta_2) \rangle_{\text{beam}} \nonumber\\
       &=\frac{224}{51821517}\alpha_A \alpha_{\bar{A}}~.
    \end{align}
\end{itemize}

We illustrate the non-uniqueness of solutions with the following example. For $S=\frac{1}{2}$, invariance under the $P$ transformation (parity) implies~\cite{ATLAS2024,CMS:2024pts,Pei:2025non}
\begin{align}
    \alpha_{\frac{1}{2},\frac{1}{2}}=-\alpha_{-\frac{1}{2},-\frac{1}{2}}~.
\end{align}
For $b=\frac{1}{2}$, we therefore obtain an alternative solution, independent of $\cos\theta_A$, that enables the extraction of $\alpha_{A/\bar{A}}$:
\begin{align}
    &\langle \hat{e}_B\cdot \hat{e}_{\bar{B}}\rangle\nonumber\\
    =&-\frac{1}{9}\alpha_A\alpha_{\bar{A}}\left(\left|\alpha_{\frac{1}{2},\frac{1}{2}}\right|^2+\left|\alpha_{-\frac{1}{2},-\frac{1}{2}}\right|^2-\left|\alpha_{\frac{1}{2},-\frac{1}{2}}\right|^2-\left|\alpha_{-\frac{1}{2},\frac{1}{2}}\right|^2 \right. \nonumber\\
   & \left. + 2 \alpha_{\frac{1}{2},\frac{1}{2}} \alpha_{-\frac{1}{2},-\frac{1}{2}}+2 \alpha_{-\frac{1}{2},-\frac{1}{2}} \alpha_{\frac{1}{2},\frac{1}{2}}\right) \\
   =& \frac{1}{9}\alpha_A\alpha_{\bar{A}}~.
\end{align}

%%%%%%%%%%%%%%%%%%%%%%%%%%%%%%%%%%%%%%%%
\section{Conclusion} \label{con} 
%%%%%%%%%%%%%%%%%%%%%%%%%%%%%%%%%%%%%%%%

In this work, we have systematically analyzed how quantum entanglement in a polarized particle–antiparticle pair $A\bar{A}$ can be accessed through the angular distributions of their subsequent decays into two-body final states. The formalism, applicable to arbitrary spins $S$ and $b$, reveals a clear distinction between bosonic and fermionic decays.

For $B$ as bosons, the entanglement-sensitive coefficient $\mathcal{C}(S,b)$ takes a simple, universal form that depends only on the spins $S$ and $b$, and is completely independent of the decay dynamics (encoded in the helicity amplitudes $H_{A/\bar{A}}$). This universality, exemplified by the constant result $\mathcal{C}(S,0)=1/2$ for a (pseudo-)scalar $B$ and the agreement of $\mathcal{C}(1,1)=1/8$ with established $W^+W^-$ results, makes bosonic channels particularly robust and attractive for model-independent entanglement verification in high-energy collisions.

In contrast, for fermionic decays, $\mathcal{C}(S,b)$ explicitly depends on the spin analysis powers $\alpha_{A/\bar{A}}$, introducing a sensitivity to the underlying decay mechanism. Consequently, measuring entanglement in such channels generally requires independent knowledge of $\alpha_{A/\bar{A}}$. We have shown how this can be achieved in $e^+e^-$ collisions by constructing specific angular observables that isolate $\alpha_A\alpha_{\bar{A}}$ when events are selected along the beam axis. Explicit expressions for these observables are provided for a range of half-integer spins $S$.

Our findings underscore that bosonic decays offer a cleaner and more direct route to test quantum entanglement, as they avoid the additional experimental and theoretical complexity associated with extracting decay-polarization parameters. The methods developed for fermionic decays, while more involved, provide a viable pathway for entanglement studies in a wider class of processes, such as those involving weakly decaying hyperons. This work provides a unified framework and practical recipes for probing quantum entanglement through angular distributions, with immediate applications in current and future collider experiments.

\begin{acknowledgments}

J. Pei is supported by the National Natural Science Foundation of China (Project NO. 12505121), by the Joint Fund of Henan Province Science and Technology R$\&$D Program (Project NO. 245200810077), and by the Startup Research Fund of Henan Academy of Sciences (Project NO. 20251820001).
L Wu is supported in part by the Natural Science Basic Research Program of Shaanxi, Grant No. 2024JC-YBMS-039.
TL is supported in part by the National Key Research and Development Program of China Grant No. 2020YFC2201504, by the Projects No. 11875062, No. 11947302, No. 12047503, and No. 12275333 supported by the National Natural Science Foundation of China, by the Key Research Program of the Chinese Academy of Sciences, Grant No. XDPB15, by the Scientific Instrument Developing Project of the Chinese Academy of Sciences, Grant No. YJKYYQ20190049, and by the International Partnership Program of Chinese Academy of Sciences for Grand Challenges, Grant No. 112311KYSB20210012. 

\end{acknowledgments}

\appendix

\section{$\mathcal{C}(S,b)$}\label{app1}

{
Wigner $d$-functions are defined as
\begin{align}
    &d^{S}_{k,j}(\theta)\nonumber\\
    = &\sqrt{(S+k)!} \sqrt{(S-k)!}
    \sqrt{(S+j)!} \sqrt{(S-j)!}\sum_{m=\max(j-k,0)}^{\min(S+j,S-k)}\nonumber\\
    &\frac{(-1)^{k-j+m}
    (\cos\frac{\theta}{2})^{2S+j-k-2m}
    (\sin\frac{\theta}{2})^{k-j+2m}}{
    (S+j-m)!(S-k-m)!m!(k-j+m)!}~,
\end{align}
which satisfy
\begin{align}
    d^{S}_{k,j}(\theta)
    = (-)^{k-j} d^{S}_{-k,-j}(\theta)
    = (-)^{k-j} d^{S}_{j,k}(\theta)~.\label{propt}
\end{align}
For the cases where $k = S$ and $j=\pm b \le S$, we have
\begin{equation}
    \max(j-k,0) = \min(S+j,S-k) = 0~.
\end{equation}
Thus, we obtain
\begin{align}
    &d^{S}_{S,\pm b}(\theta)\nonumber\\
    =&
    \frac{\sqrt{(2S)!} }{\sqrt{(S+b)!} \sqrt{(S-b)!}}
    \left(\cos\frac{\theta}{2}\right)^{S\pm b}
    \left(-\sin\frac{\theta}{2}\right)^{S\mp b}~.
\end{align}
More explicitly, we find
\begin{align}
    d^{S}_{S,b}(\theta)&=A_{S,b}(\theta)\left(\cos\frac{\theta}{2}\right)^{2 b}~, \\
    d^{S}_{S,-b}(\theta)&=A_{S,b}(\theta)\left(-\sin\frac{\theta}{2}\right)^{2 b}~,
\end{align}
where
\begin{align}
    A_{S,b}(\theta)&= \frac{\sqrt{(2S)!} \left(-\sin\theta\right)^{S-b}}{2^{S-b} \sqrt{(S+b)!} \sqrt{(S-b)!}}~.
\end{align}
The expression for $\langle \cos (2S(\phi_1-\phi_2)) \rangle$ is given by
\begin{align}
    &\langle \cos (2S(\phi_1-\phi_2)) \rangle\nonumber\\
    =&\int_{-1}^1 d\cos\theta_1 \int_{-1}^1 d\cos\theta_2 \int_0^{2\pi} d\phi_1 \int_0^{2\pi} d\phi_2~ \mathcal{W}(\theta_1,\theta_2,\phi_1,\phi_2)\nonumber\\
    &\times \frac{1}{2}\left(e^{i2S\phi_1}e^{-i2S\phi_2}+e^{-i2S\phi_1}e^{i2S\phi_2}\right)\nonumber\\
    =&\frac{(2S+1)^2}{32}\int_{-1}^1 d\cos\theta_1 \int_{-1}^1 d\cos\theta_2\nonumber\\
    &\times\sum_{\lambda_1,\lambda_{2}=\pm b}\left(1-{\rm Sign}[\lambda_1]\alpha_A\right)\left(1-{\rm Sign}[\lambda_{2}]\alpha_{\bar{A}}\right) \nonumber\\
        &\times\sum_{k,j,m,n=-S}^S \alpha_{k,j}\alpha_{m,n}^* \left(\delta_{k+2S,m}\delta_{j+2S,n}+\delta_{k-2S,m}\delta_{j-2S,n}\right)  \nonumber\\
        & d^{S}_{k,\lambda_1}(\theta_1)
        d^{S}_{m,\lambda_1}(\theta_1)
        d^{S}_{\lambda_{2},j}(\pi-\theta_2)
        d^{S}_{\lambda_{2},n}(\pi-\theta_2)~.
\end{align}
Since
\begin{align}
&\sum_{k,j,m,n=-S}^S \alpha_{k,j}\alpha_{m,n}^* \left(\delta_{k+2S,m}\delta_{j+2S,n}+\delta_{k-2S,m}\delta_{j-2S,n}\right)\nonumber\\
=& \left(\alpha_{-S,-S}\alpha^*_{S,S}+\alpha^*_{-S,-S}\alpha_{S,S}\right)~,
\end{align}
we obtain
\begin{align}
    &\mathcal{C}(S,b)= \nonumber\\
       =&\frac{(2S+1)^2}{32}\int_{-1}^1 d\cos\theta_1 \int_{-1}^1 d\cos\theta_2\nonumber\\
    &\times\sum_{\lambda_1,\lambda_{2}=\pm b}\left(1-{\rm Sign}[\lambda_1]\alpha_A\right)\left(1-{\rm Sign}[\lambda_{2}]\alpha_{\bar{A}}\right)(-1)^{\lambda_2-\lambda_1-2S} \nonumber\\
        & d^{S}_{S,\lambda_1}(\theta_1)
        d^{S}_{S,-\lambda_1}(\theta_1)
        d^{S}_{S,\lambda_{2}}(\theta_2)
        d^{S}_{S,-\lambda_{2}}(\theta_2)~,
\end{align}
where we have used Eq.~(\ref{propt}) and 
\begin{align}
    \int_{-1}^1 d\cos\theta~f(\pi-\theta)=\int_{-1}^1 d\cos\theta~f(\theta)~.
\end{align}
Given that $\lambda_{1/2}=\pm b$, we have
\begin{align}
   & d^{S}_{S,\lambda_{1/2}}(\theta_{1/2})
        d^{S}_{S,-\lambda_{1/2}}(\theta_{1/2})=\frac{{(2S)!} \left(-\sin\theta_{1/2}\right)^{2S}}{2^{2S} {(S+b)!} {(S-b)!}}~.
\end{align}
Therefore,
\begin{align}
    &\mathcal{C}(S,b)\nonumber\\
       =&\frac{(2S+1)^2}{32}\int_{-1}^1 d\cos\theta_1 \int_{-1}^1 d\cos\theta_2 \sum_{\lambda_1,\lambda_{2}=\pm b}\nonumber\\
    &\left(1-{\rm Sign}[\lambda_1]\alpha_A\right)\left(1-{\rm Sign}[\lambda_{2}]\alpha_{\bar{A}}\right)(-1)^{\lambda_2-\lambda_1+2S} \nonumber\\
        & \left(\frac{{(2S)!} }{ {(S+b)!} {(S-b)!}}\right)^2
        \frac{1}{2^{4S} } \left(\sin\theta_1\right)^{2S}\left(\sin\theta_2\right)^{2S}~.
\end{align}
Additionally, we have
\begin{align}
    \int_{-1}^1 d\cos\theta ~\left(\sin\theta\right)^{2S}=\sqrt{\pi}\frac{\Gamma(S+1)}{\Gamma(S+\frac{3}{2})}
\end{align}
and
\begin{align}
   & \sum_{\lambda_1,\lambda_{2}=\pm b}\left(1-{\rm Sign}[\lambda_1]\alpha_A\right)\left(1-{\rm Sign}[\lambda_{2}]\alpha_{\bar{A}}\right)(-1)^{\lambda_2-\lambda_1+2S}\nonumber\\
   =& (-1)^{2(S-b)}\left(1+(-1)^{2b}\right)^2\nonumber\\
   &+(-1)^{2(S-b)}\left(-1+(-1)^{4b}\right)\alpha_A \nonumber\\
   &+(-1)^{1+2(S-b)}\left(-1+(-1)^{4b}\right)\alpha_{\bar{A}} \nonumber\\
   &+(-1)^{1+2(S-b)}\left(-1+(-1)^{2b}\right)^2 \alpha_A\alpha_{\bar{A}} \nonumber\\
   =& \begin{cases}
       4~,&\quad b=0,1,2,...\\
       -4\alpha_A\alpha_{\bar{A}}~,&\quad b=\frac{1}{2},\frac{3}{2},\frac{5}{2},...
   \end{cases}~,
\end{align}
where we have made use of the fact that $S-b$ is an integer and $4b$ is even.
Finally, we obtain
\begin{align}
  &\mathcal{C}(S,b)\nonumber\\
= &\frac{(2S+1)^2}{32} \left(\frac{{(2S)!} }{ {(S+b)!} {(S-b)!}}\right)^2
         \frac{\pi }{2^{4S} } \left(\frac{\Gamma(S+1)}{\Gamma(S+\frac{3}{2})}\right)^2 \nonumber\\
        & \times \begin{cases}
       4~,&\quad b=0,1,2,...\\
       -4\alpha_A\alpha_{\bar{A}}~,&\quad b=\frac{1}{2},\frac{3}{2},\frac{5}{2},...
   \end{cases}\nonumber\\
   =&\frac{1}{2}\frac{\Gamma^4(S+1)}{\Gamma^2(S-b+1)\Gamma^2(S+b+1)} 
   \times \begin{cases}
       1~,&\quad b=0,1,2,...\\
       -\alpha_A\alpha_{\bar{A}}~,&\quad b=\frac{1}{2},\frac{3}{2},\frac{5}{2},...
   \end{cases}~.
\end{align}
}

\nocite{*}

\bibliographystyle{jhep}
\bibliography{spinS}

\end{document}